\pdfoutput=1
\documentclass[journal]{IEEEtran}

\usepackage[T1]{fontenc}
\usepackage[utf8]{inputenc}
\usepackage{cite}
\usepackage{graphicx}
\usepackage{booktabs}
\usepackage{amsmath,amssymb,amsthm}
\usepackage{mathtools}
\usepackage{algorithm}
\usepackage{algorithmic}
\usepackage{microtype}
\usepackage{xspace}
\usepackage{url}
\usepackage{xcolor}
\usepackage{hyperref}
\hypersetup{hidelinks}

\newtheorem{theorem}{Theorem}
\newtheorem{lemma}{Lemma}
\newtheorem{corollary}{Corollary}
\newtheorem{definition}{Definition}
\newtheorem{assumption}{Assumption}
\newtheorem{proposition}{Proposition}
\theoremstyle{remark}
\newtheorem{remark}{Remark}
\theoremstyle{plain}

\newcommand{\pk}{\mathsf{pk}}
\newcommand{\sk}{\mathsf{sk}}
\newcommand{\pp}{\mathsf{pp}}
\newcommand{\Setup}{\mathsf{Setup}}
\newcommand{\Prove}{\mathsf{Prove}}
\newcommand{\Verify}{\mathsf{Verify}}
\newcommand{\Sign}{\mathsf{Sign}}
\newcommand{\SVer}{\mathsf{SigVer}}
\newcommand{\HB}{\mathsf{HB}}
\newcommand{\supp}[1]{Supplementary Material, Section~#1}

\begin{document}

\title{Beyond Identity Creation: Bilinear Sybil Cost in Identities and Time}

\author{Homayoun~Maleki,
        Nekane~Sainz,
        Jon~Legarda,
        and~Igor~Santos-Grueiro%
\thanks{H. Maleki, N. Sainz and J. Legarda are with DeustoTech, University of Deusto, Bilbao, Spain (e-mail: h.maleki@deusto.es; nekane.sainz@deusto.es; jlegarda@deusto.es).}%
\thanks{I. Santos-Grueiro is with the International University of La Rioja (UNIR), Logro\~{n}o, Spain (e-mail: igor.santosgrueiro@unir.net).}}

\markboth{Preprint}{Maleki \MakeLowercase{\textit{et al.}}: Beyond Identity Creation}

\maketitle
\begingroup\renewcommand\thefootnote{}\footnotetext{This work has been submitted to the IEEE for possible publication. Copyright may be transferred without notice, after which this version may no longer be accessible.}\addtocounter{footnote}{-1}\endgroup

\begin{abstract}
Sybil defenses price the creation of an identity, yet influence goes, again and again, to active identities. We study the cost of maintaining influence. A defense has a bilinear Sybil cost with ratio $A$ if keeping $s$ identities active for $T$ periods costs at least $sT/A$ honest units: the average cost per identity-window is at least $1/A$ times an honest participant's cost. Recurrence alone does not guarantee this cost. The cost can be carried over from earlier periods, shared among many identities, or be negligible when fast hardware does the work of thousands of honest devices. We give four sufficient conditions for a defense that checks every identity each period to achieve bilinear Sybil cost: identity binding, freshness, a shared deadline, and sequential, non-batchable work. They yield two guarantees: a bilinear total cost, with $A$ the speed advantage of the adversary's fastest hardware, and, because the work must be fresh and finished by a common deadline, a bound on the identities active in a period by the processors running at that time. We prove that none of the four follows from the other three. A verifiable delay function on an identity-bound input meets all four without payments, with $A$ of 2 to 8 for CPUs and about 28 for a hypothetical ASIC, against $10^3$ to $10^7$ for hash puzzles. We measure one active identity per core per period against about $80{,}000$ under a signed heartbeat, no gain from batching up to $1{,}000$ identities, and rejection of all tested attacks.
\end{abstract}


\section{Introduction}\label{sec:intro}

\IEEEPARstart{A}{} Sybil attacker presents many identities to gain influence that should belong to many
participants~\cite{douceur2002sybil}. Most defenses make identities costly to \emph{create}: a CAPTCHA, a personhood
check, a phone number, an IP address, a puzzle solved once for a node
identifier~\cite{vonahn2003captcha,borge2017proofofpersonhood,baumgart2007skademlia}. Influence, however, is
allocated again and again to the identities that are active. A committee is drawn each epoch from the participants
who are currently present; a peer-to-peer network fills its routing tables with the nodes that are online, and an
attacker who controls enough of them can eclipse honest nodes~\cite{marcus2018eclipse}; a free tier renews the
allowance of every active account. A creation cost does not recur: an identity created for $K$ and kept for $T$
periods costs only $K/T$ per period, so a long enough horizon defeats any creation price, and a rate limit on creation
bounds how fast identities appear, not how many are active at once. The cost that matters is the cost of
\emph{maintaining} identities, as the distinction between stock and flow costs of attack in blockchains
suggests~\cite{budish2018limits}. A defense imposes a \emph{bilinear Sybil cost} if keeping $s$ identities active for
$T$ periods consumes at least a fixed amount of resource per identity and period: every active identity pays its own
cost, in every period, and no identity rides on the payment of another or on its own past payments.

\emph{Recurrence alone does not guarantee this cost.} Recurring defenses already charge identities again and
again~\cite{li2012sybilcontrol,gupta2018proof,idena}, yet the cost of maintaining influence can still fail to grow
with $s$ and $T$. It fails in one of three ways. The cost is \emph{carried over}: phone numbers and IP addresses are paid for once and reused in
later periods. The cost is \emph{shared}: one resource or one computation serves many identities; our earlier work
showed that the cost of a resource required per identity can be undercut in this way~\cite{maleki2026ia}, but
deliberately left time out. Or the unit is \emph{negligible}: a signed heartbeat recurs and is tied to each key, yet
one laptop core keeps about $80{,}000$ identities active in a two-second window, and hash puzzles let one GPU do the
work of thousands of honest devices (Table~\ref{tab:compare}). Blockchains are no exception: their peer-to-peer layer
chooses peers among addresses and identifiers obtained once, which eclipse attacks on Bitcoin and Ethereum
exploit~\cite{heilman2015eclipse,marcus2018eclipse}, and their consensus resources fall into the same failures
(Section~\ref{sec:discussion}). A per-period fee avoids all three, but needs a currency and a payment system, ties
identities to financial accounts, is unavailable in many open systems, and bounds only the total cost, not the
resources an adversary runs at once. We ask:

\begin{quote}
\emph{Under what conditions does keeping $s$ identities active for $T$ periods cost an adversary at least in proportion to
$sT$, and can these conditions be met without payments?}
\end{quote}

We give sufficient conditions and show that they can be met. In every period, each identity must deliver the output
of a verifiable delay function (VDF)~\cite{boneh2018vdf} on an input derived from its key, the period number and a
fresh beacon value, before a deadline that is the same for all identities. A VDF takes a minimum time of sequential
computation that parallel hardware cannot shorten, and is verified in milliseconds. The input is fixed only when the
period starts, so the work cannot be carried over; it depends on the key, so it cannot be shared; and it is
sequential, so faster hardware gains only a small factor $A$, 2 to 8 for CPUs and about 28 for
a hypothetical ASIC, against $10^3$ to $10^7$ for hash puzzles. Because the deadline is shared, an attacker that wants $s$ identities active needs about
$s/A$ processors running in the same period, however many keys it registers.

\textbf{Contributions.} \emph{(1) A cost model} (Section~\ref{sec:model}). We separate the maintenance cost $C(s,T)$
from creation cost, in units of any resource, define bilinear Sybil cost, $C(s,T) \ge sT/A$, and identify the three
ways a recurring cost fails it: carried over, shared and negligible.
\emph{(2) Sufficient conditions and two guarantees} (Sections~\ref{sec:bpc}--\ref{sec:necessity}). We give four
sufficient conditions for bilinear Sybil cost, stated for defenses that require a proof from every identity in every
window (Definition~\ref{def:bpc}): identity binding (C1), freshness (C2), a shared deadline (C3), and
sequential, non-batchable work (C4), and prove that together they give two guarantees: the total cost is bilinear,
$C(s,T) \ge sT/A$, and an adversary with $s$ identities active in a window runs about $s/A$ processors during that
window, however many keys it has registered (Lemma~\ref{lem:window}, Theorem~\ref{thm:main}). C1 and C4 carry the
total cost; C2 and C3 place it inside one common slot per window, which yields the bound on simultaneous hardware.
None of the four follows from the other three (Theorem~\ref{thm:independence}): without C1 the cost no longer scales
with $sT$, without C4 its ratio $A$ is no longer small, and without C2 or C3 the bound on simultaneous hardware is
lost. The conditions are sufficient; we do not claim that they are necessary for every possible defense.
\emph{(3) A realization without payments, and its evaluation} (Sections~\ref{sec:construction}
and~\ref{sec:eval}). A VDF over class groups, a signature scheme, a beacon and a log, in which every identity and
window evaluates in its own class group, so that no batching holds in the generic group model; and measurements on two
machines of every assumption and prediction.

\textbf{A cost in processor time.} We state the guarantee in resource units and leave money out on purpose. A bound in
money rests on prices and on how efficiently hardware turns money into work, and both change; processor time is
already cheap, below one cent per identity and month at current cloud prices. The bound in resource units does not
depend on prices, and on hardware only through $A$. What an adversary cannot buy is time inside a slot: $s$ identities
active in the same slot need about $s/A$ processors running at the same time, which is what attacks that need many
identities at once, such as filling a committee or eclipsing a node, require. Our scheme does not replace
creation-time defenses such as a CAPTCHA or a phone number but complements them: those decide which keys are
registered, and ours limits how many registered keys an adversary can keep active, about $A$ per processor in each
window.

\section{Related Work}\label{sec:related}

\textbf{Admission-time defenses.} Without a trusted authority a single entity can present many
identities~\cite{douceur2002sybil}. Most practical defenses charge when an identity is created or admitted: CAPTCHAs
require a human answer~\cite{vonahn2003captcha}; pseudonym parties require physical
presence~\cite{ford2008pseudonym,borge2017proofofpersonhood}; social-graph defenses bound the identities an attacker
attaches to honest users~\cite{yu2006sybilguard,yu2008sybillimit}; anonymous tokens rate-limit what a client
obtains~\cite{davidson2018privacypass}. Such a cost, once paid, covers every later use of the identity.

\textbf{Recurring and resource-based defenses.} Pricing via processing~\cite{dwork1993pricing} and proof of
work~\cite{nakamoto2008bitcoin} tie influence to computation. SybilControl~\cite{li2012sybilcontrol} asks nodes to
solve puzzles periodically, and RTCT~\cite{parkhomenko2026rtct} issues hash puzzles at random times. Resource-burning
defenses make honest participants spend less than an attacker as the attack grows~\cite{gupta2018proof,gupta2021ergo},
at the price of estimating the attack; we instead fix a small honest cost, and the bound holds whatever the size of
the attack. Idena~\cite{idena} renews identities each epoch in synchronous sessions. The computational defenses rely on
hash puzzles, which are parallelizable, so one GPU or ASIC does the work of thousands to millions of honest devices
(Table~\ref{tab:A}); memory-hard functions~\cite{biryukov2016argon2} narrow this gap but remain parallelizable.

\textbf{Sequential work and VDFs.} Time-lock puzzles~\cite{rivest1996timelock} and proofs of sequential
work~\cite{mahmoody2013posw,cohen2018simple} require computation that cannot be parallelized; VDFs add a unique output
and fast verification~\cite{boneh2018vdf,wesolowski2019efficient,pietrzak2019simple}. Their proofs of replication
already bind the VDF input to a unique identifier~\cite{boneh2018vdf}. FCLA binds a VDF to a public challenge and to
auditable receipt records but considers neither identities nor many of
them~\cite{yesmukhanov2026fcla}; VDF challenges have been combined with classification to detect Sybil nodes in
vehicular networks~\cite{hadri2026vanet}, without a cost model over many processors. Batch solving of time-lock puzzles
shows that the cost of many inputs need not add up~\cite{srinivasan2023batchable,dujmovic2024batch}, which is why we
state and prove no batching explicitly (Section~\ref{sec:construction}).

\begin{table*}[t]
\centering
\caption{Defenses in the terms of this paper: the failure each falls into, and the condition of Section~\ref{sec:bpc}
it does not meet.}\label{tab:compare}
\small
\begin{tabular}{@{}lll@{}}
\toprule
Defense & Failure & Condition not met \\
\midrule
CAPTCHA~\cite{vonahn2003captcha}, phone number, social graph~\cite{yu2006sybilguard} & carried over & C1: work not tied to the window \\
Limit per IP address (e.g., Tor relays~\cite{maleki2026ia}) & carried over (bounded sharing) & C1: one address, several identities \\
Puzzle for a node identifier~\cite{baumgart2007skademlia} & carried over & C1: solved once \\
Signed heartbeat & negligible & C4: no costly work \\
Periodic hash puzzles~\cite{li2012sybilcontrol,parkhomenko2026rtct} & negligible ($A = 10^3$--$10^7$) & C4 only with large $A$; C3 \\
Resource burning~\cite{gupta2018proof,gupta2021ergo} & negligible ($A = 10^3$--$10^7$) & C4 only with large $A$ \\
Shared sequential computation~\cite{spacemeshpoet} & shared & C1: work not tied to the identity \\
Bitcoin, Ethereum peer tables~\cite{heilman2015eclipse,marcus2018eclipse} & carried over & C1: addresses, node IDs kept \\
Proof of work~\cite{nakamoto2008bitcoin} & negligible ($A \approx 6.5\times10^7$) & C4 only with large $A$ \\
Proof of stake~\cite{kiayias2017ouroboros} & carried over (in units) & C1: same stake in every epoch \\
PoVF~\cite{xiong2025povf}, Delay Towers~\cite{motepalli2022delay} & concurrency bound not implied & C2, C3 \\
Pseudonym parties~\cite{ford2008pseudonym}, Idena~\cite{idena} & human resource & outside the computational C4 \\
\textbf{Identity-bound VDF scheme (Sec.~\ref{sec:construction})} & \textbf{none} & \textbf{C1--C4 met (Thm.~\ref{thm:construction})} \\
\bottomrule
\end{tabular}
\end{table*}

\textbf{Identities bounded by hardware.} Proofs of space bound an adversary's fraction of nodes in a distributed hash
table by its fraction of disk space~\cite{gunther2025sybils}. With sequential work, Delay Towers bind chains of VDF
proofs to a node's key~\cite{motepalli2022delay}, and PoVF requires each node to keep computing VDF outputs before a
deadline, arguing informally that a node then maintains at most as many identities as it has
cores~\cite{xiong2025povf}. In both, each node extends its own chain, whose next input is its previous output: there is
no fresh public value (C2), and each node has its own timeout or quota rather than a shared deadline (C3). By
Section~\ref{sec:necessity}, this lets the work be spread out in time, so the per-core bound is not implied. PoVF also
evaluates all chains in one RSA group fixed at genesis, whose trapdoor shortcuts every chain; we found no analysis of
batching or of the speed ratio in either design, and both ask honest nodes to compute continuously (detailed comparison in
\supp{S-V}). We give the properties under which the bound holds, show what fails without each, and need only seconds
of honest computation per period.

\section{Model}\label{sec:model}

\subsection{Time, windows and public inputs}

Time is divided into consecutive \emph{windows} $d = 1, 2, \ldots$, the periods of the introduction.
Window~$d$ contains a \emph{slot} $[a_d, b_d]$ of length $\Delta = b_d - a_d$,
and the slots of different windows do not overlap.

\textbf{Beacon.} At time $a_d$ a public randomness beacon publishes a fresh value
$\eta_d \in \{0,1\}^\lambda$, unpredictable before $a_d$
(e.g.\ a threshold beacon such as drand~\cite{syta2017drand}).

\textbf{Time evidence.} A public, append-only log records each submitted proof together with the
time it was received, and later lets anyone check whether a given proof was received by
a given time. We call the record returned by the log \emph{time evidence} and denote it by~$\rho$.

\begin{assumption}[Trusted time evidence]\label{asm:time}
The log does not backdate entries: if $\rho$ shows that a proof was received by time~$b$,
then the proof existed at time~$b$.
\end{assumption}

The trust can be spread over $n$ independent witness logs, a proof being accepted when at least $k$ of them record
receipt by the deadline; Assumption~\ref{asm:time} is then needed only for fewer than $k$ colluding witnesses. The
deadline $b_d$ is \emph{shared}: it is the same for every identity in window~$d$.

\subsection{Identities, participants and adversary}

An identity is a signature key pair $(\pk, \sk)$ whose public key is registered with the
system. The schemes we study do not limit how many keys a party registers; they control what it costs
to keep each of them \emph{active}. An identity is active in window~$d$ if an accepted proof
for $(\pk, d)$ exists. An honest participant controls one identity and runs the prover on one device.
The slowest device the system supports is the \emph{reference device}; $t_h$ is the time it needs to produce one
proof. Here the reference device is a laptop, desktop or small cloud machine (phones: Section~\ref{sec:limitations}).

The adversary controls any number of identities and any amount of hardware of any kind
(CPUs, GPUs, ASICs, rented cloud machines). It sees every public value as soon as it is
published, may start computing at any time, and may delegate computation.
It is computationally bounded with respect to the signatures and the
hash function, modeled as a random oracle~\cite{bellare1993random}.
Its goal is to keep $s$ identities active in each of $T$ consecutive windows at minimum cost.

\subsection{Maintenance cost and bilinear Sybil cost}

A defense charges identities in some resource: a fee, human labor, an address, computation. We count only the cost
of \emph{maintenance}. Creating an identity, that is, registering its key and meeting any requirement checked only at
admission, is not counted; everything else the adversary spends to keep identities active is counted, whenever it is
spent, including work done in advance. We measure cost in units of the resource; one \emph{honest unit} is what an
honest participant spends to keep its own identity active for one window.

\begin{definition}[Maintenance cost]\label{def:cost}
$C(s, T)$ is the minimum amount of the resource, beyond creation and in honest units, that any adversary must spend to
keep $s$ identities active in each of $T$ windows.
\end{definition}

A defense that charges only at creation, such as a CAPTCHA or a phone number checked once, has no maintenance cost:
nothing is spent per window, and $C(s,T) = 0$. We call $\bar e(s,T) = C(s,T)/(sT)$ the
\emph{cost per identity-window}.

\begin{definition}[Bilinear Sybil cost]\label{def:ric}
A defense has a \emph{bilinear Sybil cost} with ratio $A \ge 1$ if, for all $s, T \ge 1$,
$C(s,T) \ge sT/A$.
\end{definition}

Equivalently, $\bar e(s,T) \ge 1/A$: on average, every identity-window costs at least a $1/A$ fraction of an honest
unit, a cost neither paid once for all windows nor shared with other identities. The definition bounds the total;
under conditions C1--C4 of Section~\ref{sec:bpc}, each identity-window costs at least $1/A$ on its own (Lemma~\ref{lem:window}). The definition
captures the scaling of the cost, not its magnitude, and a defense with $C(s,T) > 0$ can fall short in either. Either
$\bar e$ has no positive lower bound; since it is positive on any bounded range of $s$ and $T$, it then becomes
arbitrarily small as $T$ grows, as $s$ grows, or as both grow together, and the cost does not scale with $sT$. Or
$\bar e$ has a positive lower bound $c$: the defense has bilinear Sybil cost with $A = 1/c$, and the question is
whether $A$ is small.
\begin{itemize}
  \item \textbf{Carried over.} For a fixed $s$, $\bar e(s,T)$ becomes arbitrarily small as $T$ grows: what keeps the identities active is
        paid once and reused. A defense that charges only at creation is the extreme case, $\bar e = 0$; a resource
        acquired once and kept, such as an IP address, and work done once for an identity and reused in every window
        (Theorem~\ref{thm:independence}(a)) behave the same way.
  \item \textbf{Shared.} For a fixed $T$, $\bar e(s,T)$ becomes arbitrarily small as $s$ grows. The typical cause is identity
        amortization~\cite{maleki2026ia}, in which one resource instance supports several identities. With bounded
        sharing, a fixed number of identities per instance, it lowers the cost by a factor of at most the error factor $E$
        of~\cite{maleki2026ia}: if the resource is provisioned again in every window, the cost stays bilinear and only
        the ratio grows, from $A$ to $AE$; if the resource is held across windows, the cost is also carried over. The
        cost is shared only when the amortization ratio tends to zero, for instance when one computation serves all
        identities in a window. Work whose cost does not add up across identities, such as batching with a factor
        $\beta$ that grows with $s$ (Corollary~\ref{cor:beta}), would have the same effect.
  \item \textbf{Negligible.} $\bar e$ has a positive lower bound, but $A$ is so large that an adversary's unit costs a
        tiny fraction of an honest one: a signature, or a hash puzzle on a GPU.
\end{itemize}
If $\bar e$ becomes small only when $s$ and $T$ grow together, the cost is carried over and shared at once. The first
two cases cover every way of missing the definition; the third meets it, but with a large $A$. A defense may show
more than one. For our construction the
resource is processor time, and one honest unit is $t_h$ seconds of one processor.

\begin{figure}[t]
\centering
\includegraphics[width=\columnwidth]{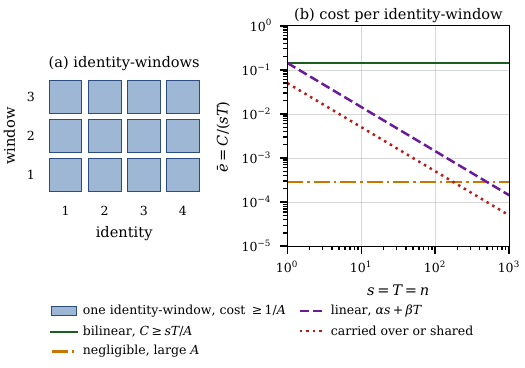}
\caption{Bilinear Sybil cost. (a) The unit of account is the identity-window, one identity during one window; each
costs at least $1/A$ honest units, so $s$ identities over $T$ windows cost at least $sT/A$. (b) Cost per
identity-window as $s$ and $T$ grow together, for illustrative constants: it stays at $1/A$ under bilinear cost, stays
positive but tiny when $A$ is large, and tends to zero for a cost linear in $s$ and $T$, which is carried over and
shared at once, and for costs that are carried over or shared.}\label{fig:bilinear}
\end{figure}

The requirement is bilinear rather than jointly linear: the bound $sT/A$ is linear in each of $s$ and $T$ when the
other is fixed, but their joint dependence is a product. A cost that is linear in $s$ and $T$ jointly, such as a price per identity plus a fixed fee
per period, $\alpha s + \beta T$, still has $\bar e \to 0$ as $s$ and $T$ grow. Bilinear Sybil cost instead takes the \emph{identity-window} as the unit of account, as the core-hour is for computation:
every pair of an identity and a window costs at least $1/A$. Sybil cost models usually count
identities~\cite{douceur2002sybil,margolin2008quantifying}, and flow-cost models of consensus count resource over
time~\cite{budish2018limits,budish2024economic}; the requirement here counts both at once.
Fig.~\ref{fig:bilinear} illustrates the difference.

\section{Four Conditions}\label{sec:bpc}

\begin{definition}[Participation proof scheme]\label{def:bpc}
A per-window participation proof scheme (a \emph{scheme}) is a tuple of algorithms $(\Setup, \Prove, \Verify)$:
$\Setup(1^\lambda) \to \pp$; $\Prove(\pp, \sk, \pk, d, \eta_d) \to \pi$, a proof for identity $\pk$ in window~$d$; and
$\Verify(\pp, \pk, d, \eta_d, \pi, \rho) \to \{0, 1\}$, which checks the proof and its time evidence.
Each proof $\pi$ has a \emph{work component} $w(\pi)$: the part whose production consumes the resource
(in Section~\ref{sec:construction}, the VDF input and output). The rest, such as a
signature, binds it to an identity and a window.
\emph{Correctness:} if an honest participant runs $\Prove$ on the reference device starting at $a_d$
and submits the result to the log, then $\Verify$ accepts.
\end{definition}

C1 keeps the cost from being carried over or shared, C4 keeps it from being negligible, and
C2 and C3 fix when it must be paid: inside one common slot per window.

\textbf{(C1) Identity binding.} Without $\sk$, no efficient adversary can produce a proof that
$\Verify$ accepts for $(\pk, d)$, except with negligible probability; and the work component of a proof accepted
for $(\pk, d)$ appears in no accepted proof for any other pair $(\pk', d')$, except with negligible probability.
\emph{Each unit of work serves one identity in one window.}

\textbf{(C2) Freshness.} No computation before $a_d$ can contribute to an accepted proof for window~$d$. Formally,
for every efficient adversary there is one that starts at $a_d$, uses the same resources after $a_d$, and, except with
negligible probability, outputs accepted proofs for window~$d$ for the same identities. In particular, no adversary that stops before $a_d$ outputs an accepted proof
for window~$d$.
\emph{Work for a window cannot be done in advance.}

\textbf{(C3) Timeliness.} $\Verify$ accepts a proof for window~$d$ only if $\rho$ shows that it
was received by $b_d$, the deadline shared by all identities.
\emph{Work for a window must be done inside one common slot.}
C3 is Douceur's requirement that resource challenges be issued to all identities simultaneously~\cite{douceur2002sybil}.

\textbf{(C4) Sequential, non-batchable cost.} There are parameters $\sigma > 0$ and $A \ge 1$ with
$\sigma = t_h / A$ such that any adversary that starts no earlier than $a_d$ and uses units $1, \ldots, p$ of a
rate-bounded resource (processors, devices, people), unit~$i$ for a total time $\tau_i$, outputs at most
$\sum_{i} \lfloor \tau_i / \sigma \rfloor$ distinct work components of accepted proofs for window~$d$; in particular,
$p$ units used for time $\tau$ each yield at most $p \lfloor \tau/\sigma \rfloor$.
\emph{Each unit of work occupies a unit of the resource for real time, which better hardware shortens only by $A$.}

A resource is \emph{rate-bounded} if each of its units performs one task at a time; we write \emph{processors} for its
units, but the bounds hold unchanged for devices or a person's attention. The theory is stated for any rate-bounded
resource; our construction and measurements use processor time only, and another resource would need its own
realization and its own measured $A$. The unavoidable cost is the time a unit is
occupied; energy and money follow from it but depend on hardware and prices. C1 says which identity and window a unit
of work serves: exactly one. C4 says what a unit costs: at least $\sigma$ of sequential computation however fast the
hardware, $A$ being the speed ratio between the fastest available hardware and the reference device, and distinct units
cannot be computed together for less than the sum of their costs. C4 is quantitative: every scheme in which each work component takes a positive minimum time and no
computation serves several components satisfies it for some $A$, and through Corollary~\ref{cor:share} the single number $A$ predicts the
largest share of active identities an attacker can hold. Resources that are held rather than occupied, such as
locked stake or storage, need their own form of C4.

\textbf{Payments and real costs.} A fee $\varphi$ paid by every identity in every window meets the definition
directly: measured in money, $C(s,T) = sT\varphi$, with $A = 1$. Bilinear Sybil cost is therefore not hard to impose in
principle; what is hard is to impose it without payments. A fee needs a currency and a payment system, ties identities
to financial accounts, and its value depends on prices. It also bounds only the total cost: an adversary can pay for
any number of identities at once, so a fee gives no bound on the resources it must run at the same time. The
construction of Section~\ref{sec:construction} instead uses a real cost, processor time, that is consumed in every
window and cannot be paid in advance or transferred, and, through the shared deadline, also bounds the processors
running at once. This makes it a building block for any mechanism that counts active participants, including
committee selection in consensus protocols.

\section{Sufficiency}\label{sec:cost}

\begin{lemma}[Per-window bound]\label{lem:window}
Let a scheme satisfy C1--C4. To keep $s$ identities active in window~$d$, an adversary must spend
at least $s\,\sigma$ seconds of processor time inside the slot $[a_d, b_d]$, and must run at least
$\lceil s / \lfloor \Delta / \sigma \rfloor \rceil$ processors during the slot.
\end{lemma}

\begin{proof}
By C2, computation before $a_d$ does not help, so we may assume that the adversary starts at $a_d$; by C3, a proof
must be received by $b_d$. All useful work for window~$d$ therefore happens inside the slot, of length $\Delta$. The adversary needs
accepted proofs for $s$ distinct identities, and by C1 their work components are pairwise distinct. Let it use
processors $1, \ldots, p$, processor~$i$ for a time $\tau_i \le \Delta$ inside the slot. By C4,
$s \le \sum_i \lfloor \tau_i/\sigma \rfloor$. Hence the processor time is
$\sum_i \tau_i \ge \sigma \sum_i \lfloor \tau_i/\sigma \rfloor \ge s\sigma$, and, since each term is at most
$\lfloor \Delta/\sigma \rfloor$, $p \ge s / \lfloor \Delta/\sigma \rfloor$. (Correctness implies $t_h \le \Delta$, so
$\lfloor \Delta/\sigma \rfloor \ge \lfloor A \rfloor \ge 1$.)
\end{proof}

\begin{theorem}[Bilinear Sybil cost]\label{thm:main}
For any scheme satisfying C1--C4 and all $s, T \ge 1$,
$C(s, T) \ge sT/A$ honest units.
\end{theorem}

\begin{proof}
Slots do not overlap. By C2, work done before $a_d$ does not help in window~$d$, and by C3 work after $b_d$ is
useless; so processor time spent in one slot cannot serve another window. Summing the bound of
Lemma~\ref{lem:window} over the $T$ windows gives at least $sT\sigma = sT\,t_h / A$ seconds of processor time,
that is, $sT / A$ honest units.
\end{proof}

Let $m = \lfloor \Delta/\sigma \rfloor$ be the number of proofs one processor can complete in a slot; when
$\Delta \approx t_h$, $m = \lfloor A \rfloor$.

\begin{corollary}[Attacker share]\label{cor:share}
If $N$ honest participants are active in a window and the attacker runs $P$ processors during its slot, the
attacker controls at most $Pm$ active identities, and its share of the active identities is at most
$Pm / (N + Pm)$, however many keys it has registered.
\end{corollary}

\begin{proof}
By C4 with $\tau \le \Delta$, $P$ processors produce at most $Pm$ distinct work components in the window, and by C1
each serves at most one identity.
\end{proof}

The share bound survives any weighting that sums per-window terms, such as the number of active windows among the last
$W$ (\supp{S-I}).
If C4 holds only up to a batching factor, that is, with $\beta \sum_i \lfloor \tau_i/\sigma \rfloor$ in place of
$\sum_i \lfloor \tau_i/\sigma \rfloor$, the bound degrades gracefully.

\begin{corollary}[Partial batching]\label{cor:beta}
For any scheme satisfying C1--C3 and C4 with factor $\beta \ge 1$, $C(s,T) \ge sT/(A\beta)$.
\end{corollary}

\begin{proof}
In the proof of Lemma~\ref{lem:window}, $s \le \beta \sum_i \lfloor \tau_i/\sigma \rfloor$ gives
$\sum_i \tau_i \ge s\sigma/\beta$; the rest of the proof of Theorem~\ref{thm:main} is unchanged.
\end{proof}

A batching attack therefore does not collapse the scheme unless $\beta$ grows with $s$.
The conditions play two roles. C1 and C4 carry the total cost: each identity and window needs its own sequential work,
which excludes the three failures. C2 and C3 fix when that work is done, inside one common slot,
and so also turn the total cost into the bound on simultaneous hardware of Lemma~\ref{lem:window} and
Corollary~\ref{cor:share}. The speed ratio $A$ divides the bound but does not change its bilinear form; what makes the
bound meaningful is that $A$ is small. The proof of Theorem~\ref{thm:main} uses all four conditions: C4 bounds only
work done from $a_d$ on, and C2 ensures that work done earlier does not help, while $C(s,T)$ counts work whenever it is
done (Definition~\ref{def:cost}). The theorem does not limit how many identities exist; it limits how many stay
active per unit of resource, and the designer sets its size through $t_h$ and the window frequency.

\section{Independence of the Conditions}\label{sec:necessity}

For each condition there is a natural scheme that meets the other three, violates it, and loses one of the two
guarantees. Without C1 the cost no longer scales with $sT$, and without C4 its ratio $A$ is no longer small; without C2 or C3 the examples below keep their total cost, by the sequentiality of their VDF rather than by
Theorem~\ref{thm:main}, but lose the bound on simultaneous hardware. Independence is not necessity: a scheme outside the per-period proof model might reach
a bilinear cost by other means. Windows start every $W = a_{d+1} - a_d \ge \Delta$.

\begin{theorem}[Independence]\label{thm:independence}
Let Assumptions~\ref{asm:time}--\ref{asm:nobatch} hold (Sections~\ref{sec:model} and~\ref{sec:construction}; the
examples below use a VDF) and let the signature scheme be EUF-CMA secure.
\begin{enumerate}
  \item[(a)] \emph{C1 does not follow from C2--C4.} There is a scheme that satisfies C2--C4 but not C1 for which
        $C(s,T) \le T/A + C_{\HB}(s,T)$, the work being shared by all identities: beyond the cost of a signed
        heartbeat $C_{\HB}$, the cost no longer grows with $s$. A scheme that reuses the work across windows, with
        $C(s,T) \le s/A + C_{\HB}(s,T)$, violates both C1 and C2.
  \item[(b)] \emph{C4 does not follow from C1--C3.} The signed heartbeat satisfies C1--C3, and C4($A$) only for
        $A \ge t_h/t_{\mathrm{sig}}$.
  \item[(c)] \emph{C2 does not follow from C1, C3 and C4.} A scheme without a beacon satisfies C1, C3 and C4 but not
        C2; given enough time before the first window, one processor keeps any number of identities active.
  \item[(d)] \emph{C3 does not follow from C1, C2 and C4.} A scheme with a deadline for each identity, all
        deadlines of window~$d$ at most $a_d + W$, satisfies C1, C2 and C4 but not C3; one processor keeps up to
        $\lfloor W/\sigma \rfloor$ identities active in every window, against $\lfloor \Delta/\sigma \rfloor$ under C3.
\end{enumerate}
\end{theorem}

\begin{table*}[t]
\centering
\caption{Independence of the conditions (Theorem~\ref{thm:independence}): dropping C1 breaks the scaling of the cost and
dropping C4 its magnitude; dropping C2 or C3 removes the bound on simultaneous hardware.}\label{tab:necessity}
\small
\begin{tabular}{@{}p{3.4cm}p{4.6cm}p{3.4cm}p{4.8cm}@{}}
\toprule
Condition dropped & Example & Failure & Adversary's cost \\
\midrule
C1 and C2: work reused across windows & VDF on $H(\pk)$, signed each window & carried over & $s/A + C_{\HB}(s,T)$ \\
C1: work tied to the identity & VDF on $H(d \| \eta_d)$, signed by each identity & shared & $T/A + C_{\HB}(s,T)$ \\
C4: costly sequential work & signed heartbeat & negligible & $sT\,t_{\mathrm{sig}}/t_h$ \\
C4: small ratio $A$ & hash puzzle (in expectation) & negligible & $sT\,r_h/R$, with $R/r_h = 10^3$--$10^7$ \\
C2: freshness & VDF on $H(\pk \| d)$, no beacon & no concurrency bound & any number per processor, given lead time \\
C3: shared deadline & own deadline per identity & weaker concurrency bound & $\lfloor W/\sigma \rfloor$ per processor, not $\lfloor \Delta/\sigma \rfloor$ \\
\bottomrule
\end{tabular}
\end{table*}

Table~\ref{tab:necessity} lists the example for each part; the proofs are in \supp{S-II}, with part~(b) as Proposition~S1. Two of the examples matter
in practice. The signed heartbeat $\Prove_{\HB}(\sk, \pk, d, \eta_d) = \Sign(\sk, \pk \| d \| \eta_d)$ meets C1--C3,
but a work component then costs one signature, $t_{\mathrm{sig}}$: with signatures of tens of microseconds, one core
keeps about $80{,}000$ identities active in a two-second window (Section~\ref{sec:eval-hb}). Proving presence does not
make an identity pay. Hash puzzles are not a formal witness, since the time to a solution is random, but in
expectation they behave as C4($A$) only for $A \ge R/r_h$, the ratio between the hash rates of the adversary's hardware
and the reference device. They keep the cost bilinear, but one GPU replaces thousands of honest participants and a
mining ASIC tens of millions (Section~\ref{sec:eval-A}): a small speed ratio is what separates sequential from
parallelizable work. Without C3, with hourly windows and a slot of a few seconds, one processor serves several hundred
times more identities than under C3.

\section{Realization with Sequential Work}\label{sec:construction}

We instantiate C4 with a VDF, obtain C1 from an identity-bound VDF input and a signature, and C2 and C3 from the
beacon and the log. We call the result the \emph{identity-bound VDF scheme}.

\subsection{Building blocks and protocol}

A VDF~\cite{boneh2018vdf} is a triple $(\mathsf{V.Setup}, \mathsf{V.Eval}, \mathsf{V.Verify})$:
$\mathsf{V.Setup}(1^\lambda, t)$ outputs parameters $\pp_V$ for delay~$t$ (here $t$ and the size of the discriminant);
$\mathsf{V.Eval}(\pp_V, x)$ outputs a value $y$ and a short proof $\pi_V$ and requires $t$ sequential steps; $\mathsf{V.Verify}$ checks them quickly. For each $x$,
only one $y$ passes verification, except with negligible probability (soundness). We use the Wesolowski
construction~\cite{wesolowski2019efficient} over class groups of imaginary quadratic fields, in which evaluation is
repeated squaring in a group of unknown order; class groups need no trusted setup, so no party knows a trapdoor. We also
use an EUF-CMA secure signature scheme (Ed25519~\cite{bernstein2012ed25519}) and a hash function $H$, modeled as a
random oracle.

The VDF input of an identity in a window is $x_{\pk,d} = H(\pk \,\|\, d \,\|\, \eta_d)$, where $\|$ is an unambiguous,
fixed-length encoding; it is fixed only when $\eta_d$ is published and is different for every identity and window. As
in \texttt{chiavdf}, $\mathsf{V.Eval}$ and $\mathsf{V.Verify}$ first derive from $x$ a negative prime discriminant
$D_x = \mathsf{DeriveDisc}(x)$ of the size fixed in $\pp_V$, and compute $y = g^{2^t}$ in the class group of $D_x$. In
\texttt{chiavdf}~1.2.0, $\mathsf{DeriveDisc}$ expands $x$ with SHA-256 and a counter into a 1024-bit integer $p$ with
$p \equiv 7 \pmod 8$, repeats until $p$ passes a primality test, and sets $D_x = -p$; the starting element is the form
$g = (2, 1, c)$ of discriminant $D_x$. Since $D_x$ has a single prime factor, the class number is odd, and squaring is a
permutation of the group. Algorithm~\ref{alg:scheme} gives the prover and the verifier.
A verifier needs only public data, so verification can be repeated by anyone, at any later time.

\begin{algorithm}[t]
\caption{Identity-bound VDF scheme}\label{alg:scheme}
\begin{algorithmic}[1]
\STATE \textbf{$\Prove(\pp, \sk, \pk, d, \eta_d)$:}
\STATE \quad $x \gets H(\pk \,\|\, d \,\|\, \eta_d)$;\; $D_x \gets \mathsf{DeriveDisc}(x)$
\STATE \quad $(y, \pi_V) \gets \mathsf{V.Eval}(\pp_V, x)$ \hfill \textit{(in the class group of $D_x$)}
\STATE \quad $\varsigma \gets \Sign(\sk, \pk \,\|\, d \,\|\, \eta_d \,\|\, y)$
\STATE \quad \textbf{return} $\pi = (\pk, d, y, \pi_V, \varsigma)$ \hfill \textit{(submitted to the log)}
\STATE \textbf{$\Verify(\pp, \pk, d, \eta_d, \pi, \rho)$:} accept iff
\STATE \quad $\eta_d$ is the beacon value of window $d$,
\STATE \quad $\mathsf{V.Verify}(\pp_V, H(\pk \,\|\, d \,\|\, \eta_d), y, \pi_V) = 1$,
\STATE \quad $\SVer(\pk, \pk \,\|\, d \,\|\, \eta_d \,\|\, y, \varsigma) = 1$, and
\STATE \quad $\rho$ shows receipt by $b_d$
\end{algorithmic}
\end{algorithm}

\subsection{Security}

C4 rests on two assumptions about the VDF. The first is standard sequentiality~\cite{boneh2018vdf}, written with $A$.

\begin{assumption}[Sequentiality]\label{asm:seq}
On an input that is unknown in advance, the fastest available hardware needs at least $\sigma = t_h / A$
seconds to compute $\mathsf{V.Eval}$, where $t_h$ is the time on the reference device.
\end{assumption}

Evaluation is $t$ sequential squarings, so $A$ is a ratio per group operation: if one operation takes $c_h$ on the
reference device and $c_A$ on the fastest hardware, then $t_h = t\,c_h$, $\sigma = t\,c_A$ and $A = c_h/c_A$.
Sequentiality concerns one input; a scheme serves many identities, so the cost of many inputs must also add up.

\begin{assumption}[No batching]\label{asm:nobatch}
Let $x_1, \ldots, x_s$ be distinct outputs of the random oracle, unknown in advance. An adversary, whatever it
computed before they became known, that then uses processor~$i$ for time $\tau_i$, $i = 1, \ldots, p$, outputs correct
VDF values for at most $\sum_i \lfloor \tau_i / \sigma \rfloor$ of them.
\end{assumption}

Assumption~\ref{asm:nobatch} is the multi-instance counterpart of sequentiality~\cite{bellare2012multiinstance}.
Three routes to batching are known, and the construction closes each. Batch solving of time-lock puzzles relies on
instances built on purpose to share structure~\cite{srinivasan2023batchable,dujmovic2024batch}; here each input is a
random-oracle output fixed only after $a_d$. Related inputs give related outputs in one group ($x_3 = x_1 x_2$ implies
$y_3 = y_1 y_2$), but random-oracle outputs cannot be made related. A trapdoor (the group order) shortcuts evaluation,
but class groups have no trusted setup. Moreover, following \texttt{chiavdf}, the challenge $H(\pk \| d \| \eta_d)$
selects the discriminant, so each identity and window evaluates in its own class group. This makes the
multi-instance statement provable in the generic group model for groups of unknown order~\cite{damgard2002generic}.

\begin{proposition}[No batching in the generic group model]\label{prop:nobatch}
Let the challenges of $s$ identity-windows select pairwise distinct discriminants, and let the order of the
starting element~$g$ in each group have a prime factor $\ell > 2^\lambda$. In the generic group model, an
adversary that uses processor~$i$ for time $\tau_i$ outputs correct values in more than
$\sum_i \lfloor \tau_i/\sigma \rfloor$ of the $s$ groups only with negligible probability. Hence Assumption~\ref{asm:nobatch} holds in this model.
\end{proposition}

The proof (\supp{S-III}) applies the standard generic argument for repeated
squaring~\cite{rotem2020generically,katz2020timelock,vanbaarsen2021timelock} to each group separately: operations in
one group give no information about another, so each group needs its own chain of $t$ dependent squarings. In a single
shared group the argument would not go through. For class groups, the condition on the order is a standard heuristic,
not a theorem. Class groups are also not generic: the class number can be computed in
subexponential time $L_{|D|}[1/2]$~\cite{hafner1989rigorous}, after which one instance is fast, but it must be computed
separately for every group, hence every identity and window. The largest such computation we know of, reported in
2019, is for a 154-digit discriminant~\cite{beullens2019csifish}, and ours have about 309 digits. This is evidence that
this known attack costs far more than $\sigma$ for every group and does not batch, not a proof against all
non-generic algorithms. Against non-generic algorithms in general, Assumption~\ref{asm:nobatch} remains
an assumption, as sequentiality does; Corollary~\ref{cor:beta} gives the bound if batching were found, and
Section~\ref{sec:eval} tests the assumption.

\begin{theorem}\label{thm:construction}
If the signature scheme is EUF-CMA secure, $H$ is a random oracle, the VDF is sound, and
Assumptions~\ref{asm:time}--\ref{asm:nobatch} hold, the identity-bound VDF scheme satisfies C1--C4. Consequently, by
Theorem~\ref{thm:main}, $C(s,T) \ge sT/A$.
\end{theorem}

The proof is in \supp{S-III}: C1 and C2 follow from the signature and the beacon value in the VDF input, C3 from the
verification rule, and C4 from soundness, uniqueness and Assumptions~\ref{asm:seq} and~\ref{asm:nobatch}.

\subsection{Trust, availability and parameters}\label{sec:trust}

The beacon can be a threshold beacon such as drand~\cite{syta2017drand}; time evidence can come from $k$-of-$n$
witness logs, trusted timestamping, or a public ledger. In this construction the total cost $sT/A$ rests only on identity binding and sequential work, whenever it is done
(\supp{S-IV}), so it survives a
predictable beacon or a backdating log; such failures remove the concurrency bound, since the work can then be spread
out in time. Network delay or congestion can make an honest proof miss the deadline, which excludes it for one window
but does not lower the adversary's cost. \supp{S-IV} gives the threat model.

The delay $t$ is set so that the reference device finishes in $t_h \le \Delta - \delta_{\mathrm{net}}$, where
$\delta_{\mathrm{net}}$ covers the time to reach the log. With $\Delta = t_h + \delta_{\mathrm{net}}$, a processor
completes at most $m = \lfloor A (1 + \delta_{\mathrm{net}}/t_h) \rfloor$ proofs per slot, so network delay and clock
skew weaken the concurrency bound only by the factor $1 + \delta_{\mathrm{net}}/t_h$, 10\% for
$\delta_{\mathrm{net}} = 1$\,s and $t_h = 10$\,s. Raising $t_h$ or the window frequency raises the adversary's cost
linearly and the honest cost by the same amount. Delegation is allowed: the processor time is spent all the same,
and the signature still requires the identity's key.

\section{Validation}\label{sec:eval}

We test the two assumptions about the VDF and the predictions of Part~I on real hardware.

\subsection{Setup}

Most measurements ran on a laptop with an Intel Core Ultra~9 275HX (24 cores: 8 performance and 16 efficiency), 32\,GB
of RAM, on mains power. We repeated the single-proof and no-batching measurements on a small cloud virtual machine with
one shared vCPU at 2.0\,GHz and 1\,GB of RAM (DigitalOcean, Ubuntu 24.04). The VDF is Wesolowski over class groups
in \texttt{chiavdf}~1.2.0, used in production by the Chia blockchain~\cite{chiavdf}, with a 1024-bit discriminant;
signatures are Ed25519. The harness is in Python with one process per worker; the VDF runs in the library's C++ and
assembly code. Every VDF output and signature was verified.

\textbf{Cost of a proof.} Evaluation time grows linearly with the number of squarings, at about $140{,}000$ per second
on one performance core; everything other than the VDF takes microseconds. Verification takes 15 to 33\,ms whatever
the delay, across a factor of 86 in evaluation time, and a proof is about 100 bytes plus a key and a signature
(\supp{S-VI}).

\subsection{Assumption~\ref{asm:nobatch}: no batching}\label{sec:eval-batch}

\begin{figure}[t]
\centering
\includegraphics[width=\columnwidth]{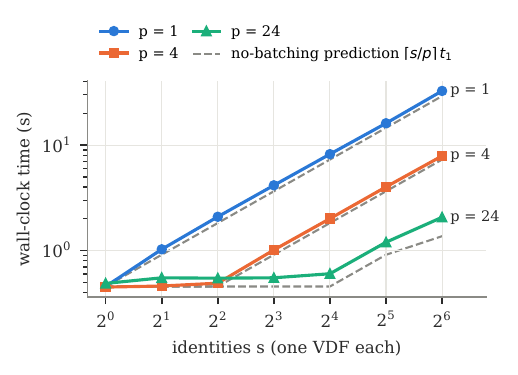}
\caption{Wall-clock time to compute VDF values for $s$ distinct identities with $p$ workers. Dashed lines:
prediction without batching, $\lceil s/p \rceil$ times one evaluation.}\label{fig:nobatch}
\end{figure}

We computed VDF values for $s$ distinct identity-bound inputs, $s \in \{1, 2, 4, \ldots, 64\}$, on
$p \in \{1, 2, 4, 8, 16, 24\}$ worker processes, with a delay of about 0.46\,s per evaluation, short enough to run all combinations. Without batching,
the wall-clock time is $\lceil s/p \rceil$ times the time of one evaluation (Fig.~\ref{fig:nobatch}).
Each combination was run three times. Across all 42 combinations, the ratio of mean measured to predicted time lies between 0.96 and 1.51 (mean 1.14), the
largest ratios only with 16 and 24 workers, on the slower efficiency cores; the processor time per identity stays
between 0.45 and 0.60\,s for every $s$. A second run with evaluations of about 0.2\,s and $s$ up to $1{,}000$ on
$p \in \{1, 8, 24\}$ gave ratios between 0.96 and 1.31: one core took 194\,s for $1{,}000$ identities, 970 times its
time for one. On the virtual machine, 1, 10 and 100 identities took 1.10 to 1.34 times the prediction.

\subsection{One identity per processor: heartbeat versus sequential work}\label{sec:eval-hb}

\begin{figure}[t]
\centering
\includegraphics[width=\columnwidth]{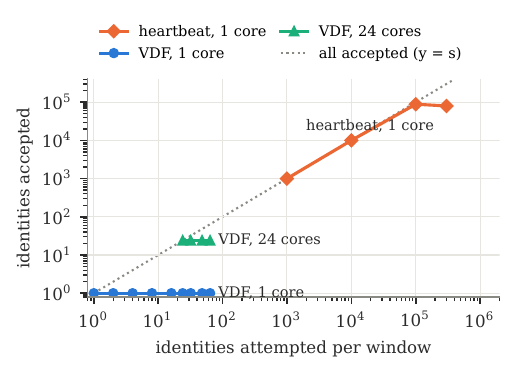}
\caption{Identities accepted in one 2-second window as a function of the identities attempted: tens of thousands per
core under the heartbeat, one per core under the identity-bound VDF scheme.}\label{fig:hb}
\end{figure}

We measured how many identities one machine keeps active in a window of $\Delta = 2$\,s, a proof counting only if it
was valid and received by the deadline. Under the signed heartbeat, one core kept up to $79{,}867$ identities active
(Fig.~\ref{fig:hb}); this is a measurement of our implementation, which signs through a Python harness, not a bound
for Ed25519, which is faster when compiled. Under the identity-bound VDF scheme, with a VDF of about 1.5\,s, chosen to
fit a 2\,s window, on 1, 8 and 24 cores and up to 64 identities, the number of accepted identities was exactly
$\min(s, p)$ in every window: one identity per core, as Lemma~\ref{lem:window} predicts for
$\lfloor \Delta/\sigma \rfloor = 1$.

\subsection{The conditions in the full protocol}\label{sec:eval-proto}

We implemented the complete protocol: a beacon, provers, a hash-chained log that records receipt times and signs its
head at the end of each window, and an independent verifier that reads only the public files and runs
Algorithm~\ref{alg:scheme} on every entry. Windows had a 2\,s slot followed by a 1\,s gap. Sixteen honest identities,
each on its own core, ran for 20 windows; all 320 identity-windows were accepted, with end-to-end latency of at most
1.86\,s. We ran five attacks, five times each, alongside an honest proof (Table~\ref{tab:attacks}); the verifier
rejected all 25 attempts, each for the expected reason, and accepted every honest proof.

\begin{table}[t]
\centering
\caption{Attacks against the full protocol (5 attempts each).}\label{tab:attacks}
\small\setlength{\tabcolsep}{4pt}
\begin{tabular}{@{}llc@{}}
\toprule
Attack & Caught by & Rejected \\
\midrule
Precompute with guessed $\eta_d$ & beacon & 5/5 \\
Replay in the next window & deadline & 5/5 \\
Reuse another identity's VDF & VDF & 5/5 \\
Submit after the deadline & deadline & 5/5 \\
Signature without $\eta_d$ & signature & 5/5 \\
\bottomrule
\end{tabular}
\end{table}

\begin{figure}[t]
\centering
\includegraphics[width=\columnwidth]{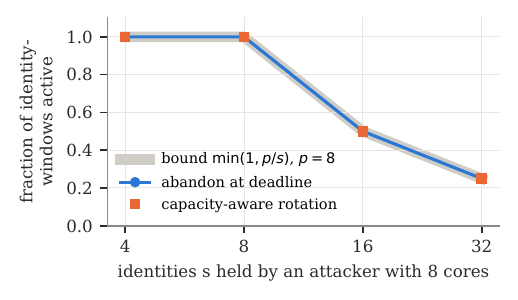}
\caption{Fraction of identity-windows an attacker with 8 cores keeps active, for two strategies, over 5, 10 and 20
windows (mean). Both are consistent with the bound $\min(1, p/s)$.}\label{fig:rational}
\end{figure}

An attacker with $p = 8$ cores tried to keep $s \in \{4, 8, 16, 32\}$ identities active for $T \in \{5, 10, 20\}$
windows, either starting all $s$ proofs and abandoning unfinished work at the deadline, or rotating $p$ proofs per window
through the identities. In all 24 runs it obtained exactly $\min(s, 8)$ accepted proofs per window, so each identity was
active in a fraction $\min(1, 8/s)$ of the windows (Fig.~\ref{fig:rational}). One window with $20{,}000$ identities was verified at about 2.9
million proofs per hour on the laptop, extrapolated from the measured rate (\supp{S-VI}).

\subsection{Assumption~\ref{asm:seq}: speed ratio}\label{sec:eval-A}

Our laptop computes about $140{,}000$ squarings per second and the virtual machine about $46{,}000$. Across 362 public
results of the \texttt{chiavdf} benchmark on x86 machines~\cite{openbenchmarking}, the slowest 2\% reach about
$54{,}000$ and the median about $174{,}000$; the fastest known CPU evaluator reached about
$360{,}000$~\cite{chiatimelord}, and a published ASIC design for class-group VDFs is 3.6 times faster than an optimized
CPU implementation~\cite{zhu2022lowlatency}. Against the fastest CPU, $A \approx 2$ with a median x86 machine as
reference device and $A \approx 8$ with our virtual machine, close to the slowest 2\%. To account for specialized
hardware, we also consider a hypothetical adversary whose hardware is 3.6 times faster than the fastest CPU evaluator,
which gives $A \approx 28$ against the virtual machine. This is an adversarial modeling assumption calibrated on the
speedup reported for the published design over its own software baseline, not the measured throughput of that design,
nor an upper bound on what specialized hardware can achieve. For hash
puzzles the same comparison gives $10^3$ to $10^7$ (Table~\ref{tab:A}).

\begin{table}[t]
\centering
\caption{Speed ratio $A$ between fast hardware and an honest core.}\label{tab:A}
\small
\begin{tabular}{@{}lll@{}}
\toprule
Work & Fast hardware & $A$ \\
\midrule
VDF (class group) & fastest CPU & $\approx 2$--$8$ \\
VDF (class group) & hypothetical ASIC & $\approx 28$ \\
SHA-256 puzzle & GPU (RTX 4090) & $\approx 3.5 \times 10^3$ \\
SHA-256 puzzle & mining ASIC & $\approx 6.5 \times 10^7$ \\
\bottomrule
\end{tabular}
\\[2pt]
\footnotesize VDF reference: measured slow machine (46k squarings/s) or median x86 (174k). SHA-256: one CPU core at
$6.1 \times 10^6$ hashes/s (measured), RTX 4090 at $2.1 \times 10^{10}$ (benchmark)~\cite{hashcat4090}, Antminer S21 at
$2 \times 10^{14}$ (vendor rating)~\cite{antminerS21}; hash ratios are estimates, not a matched workload.
\end{table}

\textbf{Money.} At the cheapest cloud spot price we found, about \$0.016 per core-hour~\cite{awsspot}, with
$t_h = 10$\,s, hourly windows and $A = 7$, the lower bound of Theorem~\ref{thm:main} is below one cent per identity and
month; but because the deadline is shared, one million identities need about $143{,}000$ cores at the same moment in
every window, and renting them only during the slots costs about six times the bound, since a common 60\,s minimum billing is paid
for 10\,s of work (\supp{S-VI}). The concurrency
requirement does not depend on prices. Under the heartbeat, one million identities need about 22\,s of one core per
window.

\section{Discussion}\label{sec:discussion}

\textbf{What the conditions predict for deployed defenses.} A defense that violates a condition is exposed in a
predictable way (Table~\ref{tab:compare}). Without C1, identities verified once by a CAPTCHA or a phone number,
relays limited per IP address, or node identifiers bought with a one-time puzzle are kept at no further cost, and an
attacker accumulates them until it fills a committee or a victim's routing table~\cite{marcus2018eclipse}. Without a
small $A$, one device replaces many honest ones. Against $N = 10{,}000$
honest participants with one device each, Corollary~\ref{cor:share} and the measured values give the attacker 89\% of
the active identities with one core under the heartbeat, 26\% with one GPU and 78\% with ten GPUs under hash puzzles
(difficulty set so that an honest core solves one per window), but 0.07\% with one core, 6.5\% with 100 cores and 41\%
with $1{,}000$ cores under the identity-bound VDF scheme with $A = 7$. With sequential work the attacker needs about
$N/A$ processors running at the same moment to match the honest participants, however many keys it registered.

\textbf{Uses.} A system that draws a committee each epoch can restrict the lottery to identities with an accepted proof
for the epoch; the attacker's expected share is then bounded by Corollary~\ref{cor:share}. A node identifier bought once
with a puzzle~\cite{baumgart2007skademlia} is kept forever; routing tables that accept only nodes with a proof for the
current window force an eclipse attacker to keep every node around its target active with its own processor time. Free tiers, faucets and periodic allowances granted per window
only to identities with an accepted proof give an attacker at most $Pm$ shares per window, not one for every account it
ever opened. How proofs are aggregated into reputation or voting weight is the application's policy; the cost guarantee
of each window holds whatever aggregation is used.

\textbf{Blockchains.} Proof of work and proof of stake weight influence by a resource rather than by identities, so the
conditions apply to them in two places. The consensus resource falls into the failures: proof of
work~\cite{nakamoto2008bitcoin} charges every period, but C4 holds only with $A \approx 6.5\times10^7$ against an
honest core, and its cost recurs in money only through energy~\cite{budish2018limits}; locked
stake~\cite{kiayias2017ouroboros} is a stock held across epochs, so in resource units its cost is carried over, as
long-range attacks with keys that held stake in the past show~\cite{deirmentzoglou2019longrange}. The network layer
does count identities: nodes choose peers among addresses and identifiers obtained once and kept, which eclipse attacks
exploit~\cite{heilman2015eclipse,marcus2018eclipse}, and admitting peers only with a proof for the current window
applies to it directly.

\textbf{From resource units to money.} An adversary that rents processors pays for their time directly. One that buys
them pays in two ways. The purchase grows with $s$, since $s$ active identities need about $s/A$ processors, but it is
paid once and carried over to every later window. The operating cost is paid in every window and is proportional to
$sT$, but it is small. The structural guarantee is that in every period $s$ active identities require about $s/A$
processors running at the same time; when such recurring resource use becomes a bilinear cost in money is an economic
question we leave to future work. For honest participants, with $t_h = 10$\,s and hourly windows, a device is busy for
under 0.3\% of the time and uses below one watt-hour per day.

\section{Limitations}\label{sec:limitations}

\textbf{Speed ratio and reference devices.} The bound is divided by $A$; the value of about 28 for specialized hardware is a
modeling assumption calibrated on a published design, not a measurement or an upper bound, and the designer can respond by raising $t_h$. We did not measure
phones, and published VDF benchmarks for mobile processors are lacking. If the reference device squares $r$ times more
slowly than ours, the designer either keeps $t_h$, so that $A$ becomes $rA$, or keeps $A$, so that honest devices
compute $r$ times longer; a phone three times slower would need 30\,s instead of 10\,s per window. Measuring $r$ and
the battery and thermal cost is future work.

\textbf{Assumptions.} No batching is proved only in the generic group model; against non-generic algorithms it rests,
like sequentiality, on the absence of known attacks, and our experiments test it on two machines. Time evidence relies
on logs that do not backdate entries; $k$-of-$n$ witnesses reduce but do not remove this trust.

\textbf{Absolute and honest cost.} The guarantee is bilinear and tunable, not large: below one cent per identity and
month in processor time, the stronger barrier being the processors needed at the same time. Honest participants pay in
every window, attack or not, unlike resource-competitive defenses~\cite{gupta2018proof,gupta2021ergo}. The scheme does
not detect Sybil identities or limit how many are registered, and a deployment needs a production beacon and log.

\section{Conclusion}\label{sec:conclusion}

Sybil defenses price the creation of identities, but influence is held over time, and its cost is the cost of
maintaining identities. We defined bilinear Sybil cost, the requirement that every active identity consume a fresh,
unshared unit of resource in every period: the cost must scale with $sT$, which fails when it is carried over or
shared, and its ratio $A$ must be small, which fails when it is negligible. Four conditions together give two complementary
results: a bilinear total cost, carried by identity binding and sequential work, and, through freshness and the shared
deadline, a bound on the identities an adversary keeps active by the processors it runs at the same time; a fee gives the first
without the second. None of the four follows from the other three.
A VDF on an identity-bound input meets all four without payments, with $A$ of 2 to 8 on CPUs and about 28 for
a hypothetical ASIC, where hash puzzles reach $10^3$ to $10^7$; on real hardware its cost added up identity by identity and
window by window. The same conditions explain where widely used defenses fall short.

\section*{Acknowledgment}
This contribution results from the Horizon Europe funded project RETIME - Urban Adaptation and Alert Solutions for a
TIMEly (Re) Action, co-funded by the European Union under Grant Agreement No. 101147113.

\bibliographystyle{IEEEtran}
\bibliography{references}

\begin{thebibliography}{10}

\bibitem{baumgart2007skademlia}
Ingmar Baumgart and Sebastian Mies.
\newblock {S/Kademlia}: A practicable approach towards secure key-based
  routing.
\newblock In {\em International Conference on Parallel and Distributed Systems
  (ICPADS)}, 2007.

\bibitem{bellare2012multiinstance}
Mihir Bellare, Thomas Ristenpart, and Stefano Tessaro.
\newblock Multi-instance security and its application to password-based
  cryptography.
\newblock In {\em Advances in Cryptology -- CRYPTO 2012}, volume 7417 of {\em
  LNCS}, pages 312--329. Springer, 2012.

\bibitem{bellare1993random}
Mihir Bellare and Phillip Rogaway.
\newblock Random oracles are practical: A paradigm for designing efficient
  protocols.
\newblock In {\em Proceedings of the 1st ACM Conference on Computer and
  Communications Security (CCS)}, pages 62--73, 1993.

\bibitem{bernstein2012ed25519}
Daniel~J. Bernstein, Niels Duif, Tanja Lange, Peter Schwabe, and Bo-Yin Yang.
\newblock High-speed high-security signatures.
\newblock {\em Journal of Cryptographic Engineering}, 2(2):77--89, 2012.

\bibitem{beullens2019csifish}
Ward Beullens, Thorsten Kleinjung, and Frederik Vercauteren.
\newblock {CSI-FiSh}: Efficient isogeny based signatures through class group
  computations.
\newblock In {\em Advances in Cryptology -- ASIACRYPT 2019, Part I}, volume
  11921 of {\em Lecture Notes in Computer Science}, pages 227--247. Springer,
  2019.

\bibitem{biryukov2016argon2}
Alex Biryukov, Daniel Dinu, and Dmitry Khovratovich.
\newblock {Argon2}: New generation of memory-hard functions for password
  hashing and other applications.
\newblock In {\em IEEE European Symposium on Security and Privacy (EuroS\&P)},
  pages 292--302, 2016.

\bibitem{boneh2018vdf}
Dan Boneh, Joseph Bonneau, Benedikt B{\"u}nz, and Ben Fisch.
\newblock Verifiable delay functions.
\newblock In {\em Advances in Cryptology -- CRYPTO 2018}, volume 10991 of {\em
  LNCS}, pages 757--788. Springer, 2018.

\bibitem{borge2017proofofpersonhood}
Maria Borge, Eleftherios Kokoris-Kogias, Philipp Jovanovic, Linus Gasser,
  Nicolas Gailly, and Bryan Ford.
\newblock Proof-of-personhood: Redemocratizing permissionless cryptocurrencies.
\newblock In {\em IEEE European Symposium on Security and Privacy Workshops
  (EuroS\&PW)}, pages 23--26, 2017.

\bibitem{budish2018limits}
Eric Budish.
\newblock The economic limits of {Bitcoin} and the blockchain.
\newblock NBER Working Paper 24717, National Bureau of Economic Research, 2018.

\bibitem{budish2024economic}
Eric Budish, Andrew Lewis-Pye, and Tim Roughgarden.
\newblock The economic limits of permissionless consensus.
\newblock In {\em Proceedings of the 25th ACM Conference on Economics and
  Computation (EC '24)}, pages 704--731. ACM, 2024.

\bibitem{chiatimelord}
{Chia Network}.
\newblock Timelords.
\newblock \url{https://github.com/Chia-Network/chia-blockchain/wiki/Timelords},
  2020.
\newblock Accessed 27 October 2026.

\bibitem{chiavdf}
{Chia Network}.
\newblock chiavdf: Verifiable delay functions over class groups.
\newblock \url{https://pypi.org/project/chiavdf/}, 2026.
\newblock Version 1.2.0.

\bibitem{cohen2018simple}
Bram Cohen and Krzysztof Pietrzak.
\newblock Simple proofs of sequential work.
\newblock In {\em Advances in Cryptology -- EUROCRYPT 2018}, volume 10821 of
  {\em LNCS}, pages 451--467. Springer, 2018.

\bibitem{damgard2002generic}
Ivan Damg{\aa}rd and Maciej Koprowski.
\newblock Generic lower bounds for root extraction and signature schemes in
  general groups.
\newblock In {\em Advances in Cryptology -- EUROCRYPT 2002}, volume 2332 of
  {\em LNCS}, pages 256--271. Springer, 2002.

\bibitem{davidson2018privacypass}
Alex Davidson, Ian Goldberg, Nick Sullivan, George Tankersley, and Filippo
  Valsorda.
\newblock Privacy pass: Bypassing internet challenges anonymously.
\newblock {\em Proceedings on Privacy Enhancing Technologies},
  2018(3):164--180, 2018.

\bibitem{deirmentzoglou2019longrange}
Evangelos Deirmentzoglou, Georgios Papakyriakopoulos, and Constantinos
  Patsakis.
\newblock A survey on long-range attacks for proof of stake protocols.
\newblock {\em IEEE Access}, 7:28712--28725, 2019.

\bibitem{douceur2002sybil}
John~R. Douceur.
\newblock The {Sybil} attack.
\newblock In {\em Peer-to-Peer Systems (IPTPS 2002)}, volume 2429 of {\em
  LNCS}, pages 251--260. Springer, 2002.

\bibitem{dujmovic2024batch}
Jesko Dujmovic, Rachit Garg, and Giulio Malavolta.
\newblock Time-lock puzzles with efficient batch solving.
\newblock In {\em Advances in Cryptology -- EUROCRYPT 2024}, LNCS. Springer,
  2024.

\bibitem{dwork1993pricing}
Cynthia Dwork and Moni Naor.
\newblock Pricing via processing or combatting junk mail.
\newblock In {\em Advances in Cryptology -- CRYPTO '92}, volume 740 of {\em
  LNCS}, pages 139--147. Springer, 1993.

\bibitem{ford2008pseudonym}
Bryan Ford and Jacob Strauss.
\newblock An offline foundation for online accountable pseudonyms.
\newblock In {\em Proceedings of the 1st Workshop on Social Network Systems
  (SocialNets)}, pages 31--36, 2008.

\bibitem{gunther2025sybils}
Christoph~U. G{\"u}nther and Krzysztof Pietrzak.
\newblock Putting {Sybils} on a diet: Securing distributed hash tables using
  proofs of space.
\newblock Cryptology ePrint Archive, Paper 2025/804, 2025.

\bibitem{gupta2018proof}
Diksha Gupta, Jared Saia, and Maxwell Young.
\newblock Proof of work without all the work.
\newblock In {\em Proceedings of the 19th International Conference on
  Distributed Computing and Networking (ICDCN)}, 2018.

\bibitem{gupta2021ergo}
Diksha Gupta, Jared Saia, and Maxwell Young.
\newblock Bankrupting {Sybil} despite churn.
\newblock In {\em IEEE 41st International Conference on Distributed Computing
  Systems (ICDCS)}, 2021.

\bibitem{hadri2026vanet}
Habiba Hadri, Mourad Ouadou, and Khalid Minaoui.
\newblock A novel approach to {Sybil} attack detection in {VANETs} using
  verifiable delay functions and hierarchical fog-cloud architecture.
\newblock {\em Journal of Cybersecurity and Privacy}, 6(2):59, 2026.

\bibitem{hafner1989rigorous}
James~L. Hafner and Kevin~S. McCurley.
\newblock A rigorous subexponential algorithm for computation of class groups.
\newblock {\em Journal of the American Mathematical Society}, 2(4):837--850,
  1989.

\bibitem{hashcat4090}
{hashcat forum}.
\newblock Benchmarks for {Nvidia}'s new 4000-series {GPUs}.
\newblock \url{https://hashcat.net/forum/post-56468.html}, 2022.
\newblock Accessed 27 October 2026.

\bibitem{heilman2015eclipse}
Ethan Heilman, Alison Kendler, Aviv Zohar, and Sharon Goldberg.
\newblock Eclipse attacks on {Bitcoin}'s peer-to-peer network.
\newblock In {\em Proceedings of the 24th USENIX Security Symposium}, pages
  129--144, 2015.

\bibitem{awsspot}
{Holori}.
\newblock {AWS} c7a.large price, us-east-1.
\newblock \url{https://calculator.holori.com/aws/ec2/c7a.large/us-east-1},
  2026.
\newblock Spot price, accessed 26 October 2026.

\bibitem{idena}
{Idena}.
\newblock Idena: Proof-of-person blockchain.
\newblock \url{https://www.idena.io}, 2026.
\newblock Accessed 27 October 2026.

\bibitem{katz2020timelock}
Jonathan Katz, Julian Loss, and Jiayu Xu.
\newblock On the security of time-lock puzzles and timed commitments.
\newblock In {\em Theory of Cryptography -- TCC 2020, Part III}, volume 12552
  of {\em LNCS}, pages 390--413. Springer, 2020.

\bibitem{kiayias2017ouroboros}
Aggelos Kiayias, Alexander Russell, Bernardo David, and Roman Oliynykov.
\newblock Ouroboros: A provably secure proof-of-stake blockchain protocol.
\newblock In {\em Advances in Cryptology -- CRYPTO 2017}, volume 10401 of {\em
  LNCS}, pages 357--388. Springer, 2017.

\bibitem{antminerS21}
{Kryptex}.
\newblock Bitmain {Antminer S21} specifications.
\newblock
  \url{https://pool.kryptex.com/es/device/asic/bitmain/antminer-s21-200th},
  2026.
\newblock Accessed 27 October 2026.

\bibitem{li2012sybilcontrol}
Frank Li, Prateek Mittal, Matthew Caesar, and Nikita Borisov.
\newblock {SybilControl}: Practical {Sybil} defense with computational puzzles.
\newblock In {\em Proceedings of the 7th ACM Workshop on Scalable Trusted
  Computing (STC)}, pages 67--78, 2012.

\bibitem{mahmoody2013posw}
Mohammad Mahmoody, Tal Moran, and Salil Vadhan.
\newblock Publicly verifiable proofs of sequential work.
\newblock In {\em Proceedings of the 4th Conference on Innovations in
  Theoretical Computer Science (ITCS)}, pages 373--388, 2013.

\bibitem{maleki2026ia}
Homayoun Maleki, Nekane Sainz, Jon Legarda, and Igor Santos-Grueiro.
\newblock Identity amortization: Shared resources and the cost of {Sybil}
  attacks.
\newblock arXiv:2610.32685, 2026.

\bibitem{marcus2018eclipse}
Yuval Marcus, Ethan Heilman, and Sharon Goldberg.
\newblock Low-resource eclipse attacks on {Ethereum}'s peer-to-peer network.
\newblock Cryptology ePrint Archive, Paper 2018/236, 2018.

\bibitem{margolin2008quantifying}
Nathan~S. Margolin and Brian~Neil Levine.
\newblock Quantifying resistance to the {Sybil} attack.
\newblock In {\em Financial Cryptography and Data Security (FC 2008)}, volume
  5143 of {\em Lecture Notes in Computer Science}, pages 1--15. Springer, 2008.

\bibitem{motepalli2022delay}
Shashank Motepalli and Hans-Arno Jacobsen.
\newblock Decentralizing permissioned blockchain with delay towers.
\newblock arXiv:2203.09714, 2022.

\bibitem{nakamoto2008bitcoin}
Satoshi Nakamoto.
\newblock Bitcoin: A peer-to-peer electronic cash system.
\newblock \url{https://bitcoin.org/bitcoin.pdf}, 2008.

\bibitem{openbenchmarking}
{OpenBenchmarking.org}.
\newblock Chia blockchain {VDF} benchmark (pts/chia-vdf).
\newblock \url{https://openbenchmarking.org/test/pts/chia-vdf}, 2026.
\newblock Accessed 27 October 2026.

\bibitem{parkhomenko2026rtct}
Ivan Parkhomenko, Larysa Myrutenko, Roman Ohiievych, and Oleksandr Laptiev.
\newblock Improving resistance to {Sybil} attacks in decentralized networks
  using the {RTCT} protocol.
\newblock In {\em Cyber Security and Data Protection (CSDP 2026)}, volume 4223
  of {\em CEUR Workshop Proceedings}, 2026.

\bibitem{pietrzak2019simple}
Krzysztof Pietrzak.
\newblock Simple verifiable delay functions.
\newblock In {\em 10th Innovations in Theoretical Computer Science Conference
  (ITCS 2019)}, volume 124 of {\em LIPIcs}, pages 60:1--60:15, 2019.

\bibitem{rivest1996timelock}
Ronald~L. Rivest, Adi Shamir, and David~A. Wagner.
\newblock Time-lock puzzles and timed-release crypto.
\newblock Technical Report MIT/LCS/TR-684, MIT Laboratory for Computer Science,
  1996.

\bibitem{rotem2020generically}
Lior Rotem and Gil Segev.
\newblock Generically speeding-up repeated squaring is equivalent to factoring:
  Sharp thresholds for all generic-ring delay functions.
\newblock In {\em Advances in Cryptology -- CRYPTO 2020}, volume 12172 of {\em
  LNCS}. Springer, 2020.
\newblock Cryptology ePrint Archive, Paper 2020/812.

\bibitem{rotem2020hidden}
Lior Rotem, Gil Segev, and Ido Shahaf.
\newblock Generic-group delay functions require hidden-order groups.
\newblock In {\em Advances in Cryptology -- EUROCRYPT 2020}, volume 12107 of
  {\em LNCS}. Springer, 2020.
\newblock Cryptology ePrint Archive, Paper 2020/225.

\bibitem{spacemeshpoet}
{Spacemesh}.
\newblock Proof of elapsed time ({PoET}).
\newblock \url{https://github.com/spacemeshos/poet}, 2026.
\newblock Accessed 27 October 2026.

\bibitem{hetznerccx}
{Spendark}.
\newblock Hetzner {CCX13} instance pricing.
\newblock \url{https://spendark.com/instances/ccx13/}, 2026.
\newblock Prices checked July 2026; accessed 27 October 2026.

\bibitem{srinivasan2023batchable}
Shravan Srinivasan, Julian Loss, Giulio Malavolta, Kartik Nayak, Charalampos
  Papamanthou, and Sri~AravindaKrishnan Thyagarajan.
\newblock Transparent batchable time-lock puzzles and applications to
  {B}yzantine consensus.
\newblock In {\em Public-Key Cryptography -- PKC 2023}, LNCS. Springer, 2023.

\bibitem{syta2017drand}
Ewa Syta, Philipp Jovanovic, Eleftherios~Kokoris Kogias, Nicolas Gailly, Linus
  Gasser, Ismail Khoffi, Michael~J. Fischer, and Bryan Ford.
\newblock Scalable bias-resistant distributed randomness.
\newblock In {\em IEEE Symposium on Security and Privacy (S\&P)}, pages
  444--460, 2017.

\bibitem{vanbaarsen2021timelock}
Aron van Baarsen and Marc Stevens.
\newblock On time-lock cryptographic assumptions in abelian hidden-order
  groups.
\newblock In {\em Advances in Cryptology -- ASIACRYPT 2021}, LNCS. Springer,
  2021.

\bibitem{vonahn2003captcha}
Luis von Ahn, Manuel Blum, Nicholas~J. Hopper, and John Langford.
\newblock {CAPTCHA}: Using hard {AI} problems for security.
\newblock In {\em Advances in Cryptology -- EUROCRYPT 2003}, volume 2656 of
  {\em LNCS}, pages 294--311. Springer, 2003.

\bibitem{wesolowski2019efficient}
Benjamin Wesolowski.
\newblock Efficient verifiable delay functions.
\newblock In {\em Advances in Cryptology -- EUROCRYPT 2019}, volume 11478 of
  {\em LNCS}, pages 379--407. Springer, 2019.

\bibitem{xiong2025povf}
Chenxi Xiong, Ting Yang, Yu~Wang, and Bing Dong.
\newblock {PoVF}: Empowering decentralized blockchain systems with verifiable
  function consensus.
\newblock arXiv:2501.01146, 2025.

\bibitem{yesmukhanov2026fcla}
Ansar Yesmukhanov and Aruzhan Tlessova.
\newblock Fresh-challenge {VDF} attestations for model-relative response
  latency.
\newblock arXiv:2609.12727, 2026.

\bibitem{yu2008sybillimit}
Haifeng Yu, Phillip~B. Gibbons, Michael Kaminsky, and Feng Xiao.
\newblock {SybilLimit}: A near-optimal social network defense against {Sybil}
  attacks.
\newblock In {\em IEEE Symposium on Security and Privacy (S\&P)}, pages 3--17,
  2008.

\bibitem{yu2006sybilguard}
Haifeng Yu, Michael Kaminsky, Phillip~B. Gibbons, and Abraham Flaxman.
\newblock {SybilGuard}: Defending against {Sybil} attacks via social networks.
\newblock In {\em Proceedings of ACM SIGCOMM}, pages 267--278, 2006.

\bibitem{zhu2022lowlatency}
Danyang Zhu, Jing Tian, Minghao Li, and Zhongfeng Wang.
\newblock Low-latency hardware architecture for {VDF} evaluation in class
  groups.
\newblock Cryptology ePrint Archive, Paper 2022/755, 2022.

\end{thebibliography}

\clearpage
\twocolumn[{\centering{\Large Supplementary Material\par}\vspace{2pt}{\normalsize Proofs, threat model, comparison with PoVF and Delay Towers, and additional measurements\par}\vspace{8pt}}]
\setcounter{section}{0}
\setcounter{table}{0}
\setcounter{figure}{0}
\setcounter{proposition}{0}
\setcounter{remark}{0}
\renewcommand{\thesection}{S-\Roman{section}}
\renewcommand{\thetable}{S-\Roman{table}}
\renewcommand{\thefigure}{S\arabic{figure}}
\renewcommand{\theproposition}{S\arabic{proposition}}
\renewcommand{\theremark}{S\arabic{remark}}
\noindent This supplement contains the proofs and details omitted from the main paper. Numbers without the prefix
``S'' (Theorem~2, Section~VII, Table~I, \dots) refer to the main paper.

\section{Composition with Additive Weighting}\label{sup:weight}

This section supports Section~V of the main paper.

\begin{remark}[Composition with additive weighting]
The bound of Corollary~1 survives any rule that gives each identity a weight equal to a sum of per-window terms, each
at most one and nonzero only in windows where the identity has an accepted proof; for example, the number of active
windows among the last $W$. If the attacker runs $P_d$ processors in window $d$, then by Corollary~1 at most $m P_d$ of
its identities are active in window $d$, so its total weight over $W$ windows is at most $m \sum_d P_d$, that is
$PmW$ for a constant $P$. When the $N$ honest participants are active in every window, the attacker's share of the
total weight is again at most $Pm/(N + Pm)$. How such weights should be chosen is a question for the system that uses
the proofs; the scheme guarantees only that every unit of weight was paid for in its window.
\end{remark}

\section{Proof of Theorem 2 (Independence)}\label{sup:indep}

We prove the four parts of Theorem~2 in turn. Windows start every $W = a_{d+1} - a_d \ge \Delta$.

\subsection{Part (a): without C1}

\textbf{Work shared by all identities.} Let $\Prove$ evaluate the VDF on $x = H(d \| \eta_d)$ and output
$(x, y, \pi_V, \Sign(\sk, \pk \| d \| \eta_d \| y))$; $\Verify$ checks the VDF on $H(d \| \eta_d)$, the signature and
the deadline. One VDF output, the work component, appears in the proofs of all identities, each of which only adds its own
signature, so C1 fails. C2 holds: the VDF input and
the signed message contain $\eta_d$, unpredictable before $a_d$, so computation before $a_d$ is independent of them and
does not help. C3 holds by the verification rule and Assumption~1. C4 holds, since an adversary that starts at $a_d$
must evaluate a distinct VDF for every distinct work component (Assumptions~2 and~3). One evaluation per window serves everyone, for $T/A$ honest units in total, plus
signing. One sequential computation serving many nodes is how some deployed systems use proofs of elapsed
time~\cite{spacemeshpoet}; for Sybil resistance the input must be derived from $\pk$. This is identity
amortization~\cite{maleki2026ia} at the level of a single computation.

\textbf{Work reused across windows.} If instead the VDF is evaluated once on $x = H(\pk)$ and the proof in every window
is $(x, y, \pi_V, \Sign(\sk, \pk \| d \| \eta_d \| y))$, the adversary evaluates $s$ VDFs once, for $s/A$ honest units,
and then signs $s$ messages per window, which is the cost of the heartbeat. This scheme violates C1, since the work
component is the same in every window, and C2, since the VDF can be evaluated before $a_d$. It is how every
admission-time defense behaves: the costly part is paid once.

\subsection{Part (b): without C4}

\textbf{The signed heartbeat.} $\Prove_{\HB}(\sk, \pk, d, \eta_d) = \Sign(\sk, \pk \,\|\, d \,\|\, \eta_d)$, accepted if
the signature verifies and $\rho$ shows receipt by $b_d$.

\begin{proposition}\label{prop:hb}
If the signature scheme is EUF-CMA secure and Assumption~1 holds, the signed heartbeat satisfies C1--C3,
and an adversary keeps $s$ identities active for $T$ windows with $C_{\HB}(s,T) \le sT\,t_{\mathrm{sig}}/t_h$, where
$t_{\mathrm{sig}}$ is the time to sign; one processor serves $\lfloor \Delta/t_{\mathrm{sig}} \rfloor$ identities per window.
\end{proposition}

\begin{proof}
C1: the work component is the signature on $\pk \| d \| \eta_d$, a message that fixes $(\pk, d)$, so one signature is
accepted for one pair only, even for an adversary that owns all the keys; without $\sk$, a proof for $(\pk, d)$ is a
forgery. C2: before $a_d$, $\eta_d$ is unpredictable, so computation before $a_d$ is
independent of the message and does not help, except by guessing $\eta_d$ (probability $2^{-\lambda}$) or by forging. C3: by the verification rule and Assumption~1. Cost: one
processor signs the $s$ messages one after another, in $s\,t_{\mathrm{sig}}$ time per window.
\end{proof}

A work component then costs $t_{\mathrm{sig}}$, so C4($A$) holds only for $A \ge t_h/t_{\mathrm{sig}}$. The bound is
bilinear, but its constant is fixed by signing speed: with signatures of tens of microseconds, one core keeps about
$80{,}000$ identities active in a two-second window (Section~VIII-C).

\textbf{Hash puzzles.} Let the work component be a solution to a hash puzzle on $H(\pk \| d \| \eta_d)$, with the
difficulty set so that the reference device, hashing at rate $r_h$, finds one in time $t_h$, and let the proof be
signed as above~\cite{li2012sybilcontrol,parkhomenko2026rtct}. C1--C3 hold as for the heartbeat, since the puzzle input
fixes $(\pk, d)$ and contains $\eta_d$. A device that hashes at rate $R$ solves $R/r_h$ puzzles in time $t_h$ in
expectation. Since the time to find a solution is random, hash puzzles do not meet the deterministic C4 and are not a
formal witness for Theorem~2(b); in expectation they behave as C4($A$) only for $A \ge R/r_h$:
about $3.5\times10^3$ for a GPU and $6.5\times10^7$ for a mining ASIC (Section~VIII-E). The bound stays bilinear but one
device replaces thousands to tens of millions of honest participants.

\subsection{Parts (c) and (d): without C2 or C3}

C2 and C3 do not change the processor time that C4 requires for each work component; they change when it must be spent.

\textbf{Without freshness.} Let the VDF input be $H(\pk \| d)$, with no beacon, and the proof be signed and checked
against the shared deadline. C1 holds, since the input fixes $(\pk, d)$; C3 holds by the verification rule; C4 holds,
since an adversary that starts at $a_d$ must still evaluate one VDF per component. C2 fails: every input is known in
advance. One processor can therefore compute the components of all identities and windows ahead of time, one after
another, and with enough lead time keep any number of identities active; the total processor time is unchanged. The same holds if the signed message also contains $\eta_d$: the proof cannot be completed before $a_d$, but its work can, so C2 in the form stated here still fails.

\textbf{Without a shared deadline.} Let the scheme be the identity-bound VDF scheme of Section~VII, except that
identity $i$ has its own deadline $b_{d,i} \le a_d + W$, where $W$ is the window period and $\Delta \le W$ the length of
the common slot in the original scheme, spread over the window period (for instance
$b_{d,i} = a_d + i\sigma$ for $i \le \lfloor W/\sigma \rfloor$). C1, C2 and C4 hold as for that scheme; C3 fails. One
processor computes the proofs one after another and keeps $\lfloor W/\sigma \rfloor$ identities active in every window;
it cannot keep more, since by C2 the work for window~$d$ starts at $a_d$, every deadline is at most $a_d + W$, and each
identity needs $\sigma$ of its time. Under C3, Lemma~1 limits it to $\lfloor \Delta/\sigma \rfloor$. With hourly
windows and a slot of a few seconds, the difference is a factor of several hundred.

With freshness and a shared deadline, $P$ processors hold at most $Pm$ active identities (Corollary~1). These two
conditions turn a bound on total work into a bound on simultaneous hardware. \hfill$\square$

\section{Security of the Construction}\label{sup:security}

\subsection{Proof of Proposition 1 (no batching in the generic group model)}

We work in the generic group model for groups of unknown order~\cite{damgard2002generic}, with one oracle per group:
each class group is accessed only through its own oracle, elements are random labels, and the group order is unknown.
A processor issues one oracle query per time step of length $c_A$.

\begin{proof}[Proof sketch]
\emph{One group.} Every element the adversary holds in a group is $g^{e}$ for an exponent it can track, and an
element obtained after a chain of $k$ dependent operations has $|e| \le 2^k$. Outputting $g^{2^t}$ after fewer than
$t$ dependent operations therefore requires an exponent $e \ne 2^t$ with $g^e = g^{2^t}$, that is, a nonzero
multiple of the order of $g$. A generic adversary learns such a multiple only from a collision between two of its
elements. With at most $q$ group operations in this group this happens with probability at most about $q^2/\ell$,
where $\ell$ is the largest prime factor of the order of $g$~\cite{damgard2002generic}; by the condition of
Proposition~1, $\ell > 2^\lambda$. Nothing is needed about the smaller prime factors. So each group needs a chain of at least $t$ dependent operations, taking time at
least $t\,c_A = \sigma$.
\emph{Many groups.} An operation in one group takes and returns elements of that group only. The groups have
independent encodings and unrelated unknown orders, so operations in other groups give no information about this one,
and the argument above applies to each group separately: each needs its own chain. A processor executes one
operation at a time, so processor~$i$, used for time $\tau_i$, completes at most $\lfloor \tau_i/\sigma \rfloor$
disjoint chains, and all processors together at most $\sum_i \lfloor \tau_i/\sigma \rfloor$. A union bound over the
$s$ groups bounds the failure probability by about $s\,q^2/2^\lambda$, negligible when $s$ and $q$ are polynomial in
$\lambda$. Two remarks on timing. Each group needs its own chain whenever it is computed, so precomputation, if the
challenges were known early, would not lower the total work. With an unpredictable beacon the challenges are unknown
before $a_d$, so the work cannot start earlier either.
Two identity-windows share a group only if their challenges yield the same discriminant, which requires a collision
of $H$ or of the discriminant derivation; the adversary cannot steer this, since $\eta_d$ is unknown before $a_d$.
\end{proof}

The single-group step is the standard generic argument for the sequentiality of repeated squaring; stronger forms
reduce speed-ups to factoring in the generic ring model~\cite{rotem2020generically} and in the strong algebraic group
model~\cite{katz2020timelock}, with extensions to general groups of unknown order~\cite{vanbaarsen2021timelock}, and
generic delay functions require hidden order~\cite{rotem2020hidden}. The step from one instance to many is what
the per-identity groups add: in a single shared group the instances would share an order, and the argument would not
go through. The Wesolowski proof costs about $t/\log t$ further operations in the same group, which only adds
per-identity work.

\subsection{Proof of Theorem 3}

\begin{proof}
\emph{C1.} The work component is $(x, y)$ with $x = H(\pk \| d \| \eta_d)$, under an unambiguous encoding, and
$\Verify$ checks $x$ against $(\pk, d)$; so the same component is accepted for another pair only if $H$ has a collision,
which in the random oracle model happens with negligible probability. This holds even for an adversary that owns all
the keys involved. In addition, an accepted proof contains a valid signature under $\pk$ on $\pk \| d \| \eta_d \| y$,
so without $\sk$ no proof for $(\pk, d)$ can be produced except by forging.
\emph{C2.} An accepted proof for window~$d$ contains a signature on a message that includes $\eta_d$ and a VDF
value on $H(\pk \| d \| \eta_d)$, a random-oracle output. Before $a_d$ the value $\eta_d$ is unpredictable, so
computation before $a_d$ is independent of every VDF input and signed message of window~$d$ and does not help, except
by guessing $\eta_d$ (probability $2^{-\lambda}$) or by forging.
\emph{C3.} By the last check of $\Verify$ in Algorithm~1 and Assumption~1.
\emph{C4.} By soundness, every accepted proof contains the correct value $y$ for its input,
and by uniqueness of the VDF output, distinct work components have distinct inputs. These inputs are random-oracle
outputs, unknown before $a_d$, so whatever the adversary computed before $a_d$, Assumption~3 bounds the number of
distinct components by $\sum_i \lfloor \tau_i / \sigma \rfloor$, where processor~$i$ is used for time $\tau_i$ after
$a_d$; Assumption~2 fixes $\sigma = t_h / A$.
\end{proof}

\section{Trust and Availability}\label{sup:trust}

This section extends Section~VII-C of the main paper.

\textbf{Realizing the services.} The beacon can be a threshold randomness beacon such as drand~\cite{syta2017drand},
whose output is unpredictable unless a threshold of its members collude. Time evidence can come from $k$-of-$n$
independent witness logs (Section~III), from standard trusted timestamping services, or from inclusion in a public
ledger whose blocks carry timestamps; each realization replaces trust in one party by trust that fewer than $k$
parties, or the ledger's consensus, misbehave. Table~\ref{tab:threat} gives the threat model.

\begin{table}[t]
\centering
\caption{Threat model for the beacon and the time evidence.}\label{tab:threat}
\small\setlength{\tabcolsep}{3pt}
\begin{tabular}{@{}p{2.6cm}p{2.5cm}p{2.7cm}@{}}
\toprule
Failure & What is lost & Mitigation \\
\midrule
Beacon predictable in advance & concurrency bound (C2) & threshold beacon; fails only if a threshold of members collude \\
Beacon late or unavailable & one window, for everyone & slot starts at recorded publication; skip the window \\
Log backdates entries & concurrency bound (C3) & $k$-of-$n$ witnesses; fails only if $k$ collude \\
Log censors honest proofs & availability of honest identities & several witnesses; tolerate missed windows \\
Network delay, congestion & availability; slack enlarges the slot & slack $\delta_{\mathrm{net}}$ (Section~VII-C) \\
\midrule
\multicolumn{3}{@{}p{7.8cm}@{}}{\footnotesize For the identity-bound VDF construction, the total cost $sT/A$ survives these failures (argued below); what can be lost is the bound on simultaneous hardware or the availability of honest identities, never the adversary's cost per identity-window.} \\
\bottomrule
\end{tabular}
\end{table}

\textbf{What a failure breaks.} The bound of Theorem~1 has two parts that depend on different services.
The \emph{total} processor time, at least $sT/A$, survives a predictable beacon and a log that backdates. This does
not follow from Theorem~1, whose C2 or C3 then fails, but from the construction: each $(\pk, d)$ still has its own
input and group, so it still needs its own chain of $t$ sequential squarings, whenever it is computed
(Proposition~1, whose argument does not depend on when the input becomes known). What such failures remove is the \emph{concurrency} bound: if $\eta_d$ is known in advance,
or the log accepts late proofs, the adversary can spread the work over time and serve many identities with one
processor (Section~VI). A beacon that publishes late shifts the slot but does not help the adversary,
since the slot is defined from the recorded publication time.

\textbf{Availability.} Network delay or congestion, including congestion caused by the adversary, can make an honest
proof miss the deadline. This excludes an honest identity for one window; it does not lower the adversary's cost. The
slack $\delta_{\mathrm{net}}$ in the choice of $t_h$, submission to several witnesses, and a policy that tolerates an
occasional missed window limit this effect; the protocol does not guarantee availability against an adversary that
controls the network. In our full-protocol runs the slowest honest proof arrived 0.14\,s before a 2\,s deadline.

\section{Comparison with PoVF and Delay Towers}\label{sup:hw}

This section extends the related work of Section~II. Table~\ref{tab:hwcompare} compares the two sequential-work designs
that bound identities by hardware with ours, in the terms of the four conditions.

\begin{table}[t]
\centering
\caption{Sequential-work designs that bound identities by hardware.}\label{tab:hwcompare}
\small\setlength{\tabcolsep}{2.5pt}
\begin{tabular}{@{}p{2.5cm}p{1.75cm}p{1.75cm}p{1.85cm}@{}}
\toprule
 & PoVF~\cite{xiong2025povf} & Delay Towers~\cite{motepalli2022delay} & Identity-bound VDF \\
\midrule
Work bound to key & seed $h(\pk \| \mathit{st})$ & key in setup & $H(\pk \| d \| \eta_d)$ \\
Fresh input (C2) & no & no & beacon \\
Shared deadline (C3) & no, per node & no, daily quota & yes \\
Group & RSA, fixed at genesis & not specified & class group per identity-window \\
Batching & not analyzed & not analyzed & Prop.~1, tested \\
Speed ratio $A$ & not analyzed & not analyzed & 2--8 (CPUs), $\approx 28$ (hypothetical ASIC) \\
Honest cost & continuous & continuous & $t_h$ per window \\
Per-core bound & argued & not analyzed & proved \\
\bottomrule
\end{tabular}
\end{table}

Both designs bind the work to a node's key, but each node extends its own chain, whose next input is the previous
output: there is no fresh public value, so the work is not fixed only when a window starts (C2), and each node has its
own timeout or epoch quota rather than a deadline shared by all (C3). By Theorem~2(c) and~(d), this is exactly what lets
the work be spread out in time, so the bound on identities per core is not implied by the design. PoVF evaluates all
chains in one RSA group whose modulus is fixed at genesis, so whoever knows its factorization evaluates every chain at
once; its argument for the per-core bound assumes that the VDF speeds up on parallel processors. We found no
analysis of batching or measurement of the speed ratio in either design, and both ask honest nodes to compute continuously. A short slot shared
by all identities forces the same concurrency while honest devices compute only seconds per period.

\section{Additional Measurements}\label{sup:meas}

This section extends Section~VIII of the main paper.

\subsection{Cost of producing and verifying a proof}

Table~\ref{tab:micro} summarizes the cost of one proof. Producing a proof is dominated by the VDF; everything else takes
microseconds. Across delays from $1.6 \times 10^4$ to $1.4 \times 10^6$ squarings (a factor of 86 in evaluation time),
the mean verification time stayed between 15 and 24\,ms. On the cloud virtual machine, verification took 43\,ms.

\begin{table}[t]
\centering
\caption{Cost of one proof on the test laptop (one core).}\label{tab:micro}
\small
\begin{tabular}{@{}lr@{}}
\toprule
Operation & Measured \\
\midrule
VDF evaluation, $1.40 \times 10^6$ squarings & $9.62 \pm 0.59$ s \\
VDF evaluation, $2.11 \times 10^5$ squarings & $1.48 \pm 0.11$ s \\
VDF verification (any delay tested) & 15--33 ms \\
VDF output and proof size & 100 B \\
Ed25519 signing / verification & 22 / 69\,$\mu$s \\
Full proof check by the verifier (Alg.~1) & 15--23 ms \\
Log entry per proof (JSON, hex-encoded) & 918 B \\
\bottomrule
\end{tabular}
\end{table}

\subsection{Heartbeat at larger scale}

Under the signed heartbeat, one core signed for up to $300{,}000$ identities per window; it kept up to $79{,}867$
identities active in one window, and $88{,}523$ in the best window at $s = 100{,}000$.

\subsection{Attackers with more identities than cores}

Fig.~4 of the main text shows the fraction of identity-windows an attacker with $p = 8$ cores keeps active, for the two
strategies of Section~VIII-D. Both are consistent with the bound $\min(1, p/s)$, and the total grew exactly as $\min(s,8)\,T$. The
cost per accepted identity-window was 1.45--1.52 core-seconds with rotation and 1.96--1.98 core-seconds when unfinished
work was abandoned. End-to-end latency of honest proofs, from the publication of $\eta_d$ to receipt by the log, was
1.65\,s on average.

\subsection{Scale of the verifier and the log}

To check the verifier and the log beyond a few dozen identities, we ran one window with $20{,}000$ identities (with a
short delay, since verification time does not depend on it). All $20{,}000$ proofs were accepted. The log appended
about $6{,}500$ entries per second in a single process, and signing the window head took 0.25\,ms. The independent
verifier checked 66 proofs per second in one process, 393 with 8 processes and 803 with 24; the laptop therefore
verifies, extrapolating from the measured rate, about 2.9 million proofs per hour, and one million proofs per hour need about five cores. Our log stores each
entry as hex-encoded JSON, about 920 bytes, or 0.9\,GB per window for one million identities; a binary encoding would be
about three times smaller, and entries need to be kept only until they have been verified, while the signed window
heads are kept permanently.

\subsection{Cost in money}

Converting Theorem~1 into money needs a price for processor time. We use the cheapest spot price of a compute-optimized
cloud instance on 26 October 2026: \$0.0313 per hour for an AWS c7a.large instance in us-east-1, whose two vCPUs are
physical cores, that is, about \$0.016 per core-hour~\cite{awsspot}; a reference device that computes for
$t_h = 10$\,s per window, hourly windows, and $A = 7$. Table~\ref{tab:money} gives the monthly cost of keeping $s$
identities active. The lower bound of the theorem counts only the processor time actually used; it is small per
identity, below one cent per month. Because the deadline is shared, however, the adversary also needs $s/A$ cores at
the same moment in every window: about $143{,}000$ cores for one million identities. Renting them only during the
slots, with the common 60\,s minimum billing, costs about six times the lower bound; keeping them all month costs over
\$1.6 million. The real cost of a large attack lies between these values. At on-demand prices (about \$0.035 to \$0.051
per core-hour on the providers we checked~\cite{awsspot,hetznerccx}) all values scale by a factor of 2.2 to 3.3. With
per-second billing and no minimum, renting only during the slots approaches the lower bound, and with hourly billing it
approaches the full-rental cost. The concurrency requirement does not depend on prices.

\begin{table}[t]
\centering
\caption{Monthly cost of keeping $s$ identities active (hourly windows, $t_h = 10$\,s, $A = 7$, \$0.016 per core-hour).}\label{tab:money}
\small\setlength{\tabcolsep}{4pt}
\begin{tabular}{@{}rrrrr@{}}
\toprule
$s$ & Bound & Cores at once & Slot rent & Full rent \\
\midrule
$10^3$ & \$4.5 & 143 & \$27 & \$1.6k \\
$10^5$ & \$453 & 14{,}286 & \$2.7k & \$163k \\
$10^6$ & \$4.5k & 142{,}857 & \$27k & \$1.63M \\
\bottomrule
\end{tabular}
\end{table}

\end{document}